\documentclass[aps,prc,twocolumn,showpacs]{revtex4}
\usepackage{graphicx,epsfig}

\newcommand{ \be }{\begin{equation}}
\newcommand{ \ee }{\end{equation}}
\newcommand{ \bea }{\begin{eqnarray}}
\newcommand{ \eea }{\end{eqnarray}}

\usepackage{color}

\begin{document}

\title{
Polarized secondary particles in unpolarized high energy 
hadron-hadron collisions?
}

\author{Sergei A.~Voloshin}

\affiliation{Department of Physics and Astronomy, 
Wayne State University, Detroit, Michigan 48201}

\date{\today} 

\begin{abstract}
In this short note I speculate on some consequences of the high 
energy collision picture in which the orbital angular momentum 
of the colliding hadrons can be converted
into secondary particle angular spin momentum via 
some spin-orbital interaction. 
In particular I discuss a possibility to observe a non-zero polarization 
of secondary particles
(e.g. hyperons) at midrapidity ($x_F=0$) and at low transverse momentum.
I also speculate that such effects could contribute 
to the produced particle directed and elliptic 
flow observed in relativistic nuclear collisions.
\end{abstract}


\maketitle

In a course of high energy hadron-hadron or nuclear collision 
the orbital angular momentum of colliding particles can be converted  
into the spin angular momentum of produced particles. 
In this case the produced particles would become polarized along 
the initial orbital angular momentum of the colliding particles. 
One particular mechanism of such a conversion
is considered in a recent paper~\cite{wang} discussing non-central nuclear 
collisions. I would totally concur with the results presented in this
paper.
Here, I discuss a few ideas beyond those already mentioned in~\cite{wang}.
Note, that the possibility of the transfer  of the orbital
momentum into spin  is not surprising, 
e.g. consider the $\rho$ resonance decaying into two pions, which
must have the orbital momentum equal to the spin of the $\rho$. 
The backward process should be also possible, 
and if the colliding pions have a particular direction of their 
orbital momentum, the spin of $\rho$ would point that very direction.

In this short note I would like to point out that such 
a conversion of the orbital 
momentum into spin (and, in principle, wise versa) 
can be relevant not only for $A+A$ collisions but also could lead to important
observable effects in hadron-hadron collisions. 
In particular I try to relate it to such phenomena as the hyperon polarization
in unpolarized hadron collisions and single-spin asymmetries 
in transversely polarized proton collisions.
Both these effect arguably are related to the orbital momentum
of the quark-gluon matter inside the constituent quarks~\cite{troshin,liang}.
I speculate that one can consider 'elementary' $p+p$ collision similar to
that of a non-central nuclear collision, namely,
introducing the notion of the reaction plane - 
the plane perpendicular to the orbital momentum, and considering particle 
production and their polarization relative to that 
plane (alternatively, to the direction of the orbital momentum).  
In that approach, for example, the hyperons could have non-zero polarization 
even at $x_F=0$ and/or small transverse momenta.

Recall that the hyperon polarization is usually measured with respect 
to the plane spanned by the hyperon's and projectile momenta, the
so-called production plane. 
It has been observed that the polarization strongly (almost linearly) depends 
on the fraction of the projectile longitudinal momentum carried, $x_F$,
and also strongly depends on the particle transverse momentum, exhibiting a
saturation at about $p_t \sim 1$~GeV/c.
In our picture the polarization of the produced particles
would be correlated mainly to the direction of the orbital momentum of 
the colliding hadrons.
Then, if measured with respect to the hyperon production plane, 
the polarization would be non-zero only
because the production plane  itself is correlated 
to the ``reaction plane'' (the orientation 
of the orbital momentum).
In this case, the observed dependence of the polarization on $x_F$ and $p_t$ 
is due not to the loss of 'actual' polarization, but due to the fact that
the hyperon's transverse momentum direction becomes less and less correlated 
with the ``reaction plane''.  
Such a sensitivity would be totally lost at $x_F=0$. 

In order to observe the 'actual' polarization one has to know 
the direction of the orbital angular momentum of the colliding hadrons. 
For that purpose we propose to use the azimuthal 
distribution of  particles in the forward rapidity region, similar to
the procedure used in the analysis of directed flow in nuclear collisions.
One could also think about this procedure from the point of view 
of single-spin asymmetries. 
When one selects the events with a particular particle azimuthal distribution,
suppose events with preferential particle emission along the $x$
axis, it means that in
these particular events the angular momentum of the system is preferentially 
pointing  along the $y$ axis. 
In its turn it would lead to the hyperon 
polarization along the $y$ axis and can be observed as correlation 
between hyperon polarization and the particles azimuthal distribution 
in the forward region.

The entire picture of $p+p$ collision would look very similar to 
that of the directed flow in nuclear collision. 
One could speculate even further: the directed flow of the produced particles
observed in high energy nuclear collision could have a significant contribution
from the same very physics that is responsible for the single-spin
asymmetries. The nuclear collision would be considered just as
``coherent'' superposition of elementary nucleon-nucleon collisions
with strong correlation of the angular momentum in each of them.

Also important that in both, $p+p$ and $A+A$, collisions 
the particle produced 
at midrapidity could be strongly polarized. For example, vector
resonances, would 
have their spin pointing along the initial orbital momentum.
Then, the  decay products 
of such resonances would have angular distribution $\propto \sin^2\theta$, 
where $\theta$ is the angle relative to the spin direction 
(in the resonance rest frame), and consequently
$\propto \cos(2\phi)$, where the angle $\phi$ is now the azimuthal angle 
with respect to the reaction plane, and thus would contribute to 
the elliptic flow (modulo distortions due to transformation from the 
resonance rest frame).
Such an additional contribution could probably explain the very strong
elliptic flow observed at RHIC (recall, that in transverse momentum region, 
$p_t \sim 3$~GeV/c elliptic flow at RHIC can not be explained by any 
model~\cite{highptv2}).

Finally I note that the effect of strong correlation between the
polarization of
hyperons produced in nuclear collisions may complicate the analysis
aimed on testing a possibility of the parity violation~\cite{parity} 
in such collisions. 
As hyperon
polarization would be along the total orbital momentum of the system,
and due to the parity violation in their decays, it would lead to
the preferential emission of the daughters of their decay along (or opposite)
to the system orbital momentum direction. The difference from the
effect discussed in~\cite{parity} would be only in the constant alignment
of the particle emission with the orbital momentum compared to
event-by event fluctuation in sign (parallel and anti-parallel to the
orbital momentum) in the original effect. The fluctuations in the hyperon
production could mask the real parity violation effect 
and special precautions should be
taken to avoid this problem.

\acknowledgments{
Discussions with R.~Bellwied and A.~Petrov are gratefully acknowledged. 
This work was supported in part by the
U.S. Department of Energy Grant No. DE-FG02-92ER40713.
}

\bibliographystyle{unsrt}

\begin{thebibliography}{99}

\bibitem{wang}
	Z.-T. Liang and X.-N. Wang, arXive:nucl-th/0410079, 2004.

\bibitem{troshin}
	S. M.~Troshin and N. E. Tyurin, arXive:hep-ph/0201267, 2002.

\bibitem{liang}
	Liang Zuo-tang and C. Boros, Phys. Rev. Lett. {\bf 79} (1997) 3608.


\bibitem{highptv2}
	STAR Collaboration, J. Adams et al., 
	nucl-ex/0407007, 2004.

\bibitem{parity}
	D. Kharzeev, arXive:hep-ph/0406125; S.Voloshin,
arXive:hep-ph/0406311, 2004.

\end{thebibliography}
 \end{document}